\documentclass[useAMS,usenatbib,usegraphicx,usegeometry]{mn2e}
\usepackage{graphicx}
\usepackage{amsmath, amssymb}

\newcommand{\oII}{[O\textsc{ii}]}

\newcommand{\alphalist}{\begin{list}{\bf{(\alph{counter})}}{\usecounter{counter}}}

\newcounter{counter}

\title[ETG Star Formation Histories in Different Environments]{Early-Type Galaxy Star Formation Histories in Different Environments}

\author[P.J. Fitzpatrick and G.J. Graves]{Patrick J. Fitzpatrick$^{1}$\thanks{E-mail:\hspace*{1mm}fitzppat@berkeley.edu\hspace*{1mm}(PJF); graves@astro.princeton.edu\hspace*{1mm}(GJG)} and Genevieve J. Graves$^{1,2}$\footnotemark[1]\\
$^{1}$Department of Astronomy, University of California, Berkeley, CA 94720, USA\\
$^{2}$Department of Astrophysical Sciences, Princeton University, Princeton, NJ 08544, USA}

\begin{document}

\maketitle

\label{firstpage}


\begin{abstract}

We use very high-S/N stacked spectra of $\sim$29,000 nearby quiescent
early-type galaxies (ETGs) from the Sloan Digital Sky Survey (SDSS) to
investigate variations in their star formation histories (SFHs) with
environment at fixed position along and perpendicular to the
Fundamental Plane (FP). We define three classifications of local group environment based
on the `identities' of galaxies within their dark matter halos:
central `Brightest Group Galaxies' (BGGs); Satellites; and Isolateds
(those `most massive' in a dark matter halo with no Satellites). We
find that the SFHs of quiescent ETGs are almost entirely determined by
their structural parameters $\sigma$ and $\Delta I_e$.  Any variation
with local group environment at fixed structure is only slight:
Satellites have the oldest stellar populations, 0.02 dex older than
BGGs and 0.04 dex older than Isolateds; BGGs have the highest
Fe-enrichments, 0.01 dex higher than Isolateds and 0.02 dex higher
than Satellites; there are no differences in Mg-enhancement between
BGGs, Isolateds, and Satellites.  Our observation that, to
zeroth-order, the SFHs of quiescent ETGs are fully captured by their
structures places important qualitative constraints on the degree to
which late-time evolutionary processes (those which occur after a
galaxy's initial formation and main star-forming lifetime) can alter
their SFHs/structures.

\end{abstract}


\section{Introduction}
There are reasons to expect early-type galaxy (ETG) star formation
histories (SFHs) to depend on a galaxy's environment because the
efficiency of relevant processes such as quenching by massive haloes
(see, e.g., \citealt{keres05, cattaneo08}), cooling flows
(see, e.g., Miller, Melott \& Gorman 1999), and satellite quenching (see, e.g., \citealt{gunn72, lea76, gisler76}) should vary with environment.  Thus, studies of
differences in the star formation histories of early-type galaxies with
environment can provide insight into mechanisms governing the formation
and evolution of these galaxies.

However, such studies have yielded somewhat contradictory results.  Many of them have suggested that, at a given luminosity, ETGs
in low-density environments are younger and more metal-rich than those
in clusters (e.g. \citealt{bernardi98, trager00b, poggianti01, terlevich02}; Caldwell, Rose \& Concannon 2003; \citealt{proctor04, thomas05, sanchez06, cooper10, rogers10}).  There are, however, conflicts with the above results: using a sample of ETGs
from the Sloan Digital Sky Survey (SDSS), \citet{bernardi06} found galaxies in the most dense environments to be older than their counterparts in the least dense environments by $\sim$1 Gyr, but found no significant differences in metallicity with environment.  Also using a sample of ETGs from SDSS, \citet{gallazzi06} found
evidence that ETGs in low-density environments were less metal-rich
than those in high-density environments.  \citet{harrison11},
studying a sample of ETGs drawn from four clusters, found
no significant differences in the ages, metallicities, or
$\alpha$-element abundance ratios between galaxies within clusters and
those found in their outskirts.  Comparing early-type galaxies in
clusters and their field contemporaries, \citet{rettura11} found no
difference in the ages of cluster and field ETGs, but
found that field ETGs were formed over longer time-scales
than those in clusters.  It is clear that there
is still much to be understood about the relationship between the
stellar population properties (SPPs) of early-type galaxies and their
environments.

One possible explanation for the contradictions among these results is entanglement between trends in the stellar populations with environment and those with other galaxy parameters.  For example, the above studies distinguished the
environments of ETGs by comparing galaxies in groups,
clusters, and the field.  Such an environmental distinction is made
ambiguous by known correlations between ETG
$M_{\star}$ and environment, such that both high-mass, bright red galaxies
and low-mass, faint red galaxies are preferentially found in denser
environments (e.g. \citealt{hogg03, mo04, blanton05b, croton05, hoyle05}).
If the SPPs show trends with galaxy parameters that are environmentally
dependent, such as $M_{\star}$, then studies comparing the stellar
populations of all galaxies at fixed environment vs. those comparing
galaxies at, for example, fixed $M_{\star}$ or fixed central stellar
velocity dispersion $\sigma$ and fixed environment will all give
different results.

To zeroth-order, ETGs are observed to form a one-dimensional (1D) family
with their SPPs showing strong trends with
galaxy mass.  Studies have sought to characterize these trends in the
stellar populations of ETGs along luminosity $L$, $M_{\star}$, or
$\sigma$.  In general, these studies have found stellar metallicities to
increase with increasing galaxy mass -- the well-known mass-metallicity
relation (e.g. \citealt{henry99, nelan05}; Smith, Lucey \& Hudson 2007; \citealt{koleva11}).  These
studies have also established that the SPPs of more massive ETGs tend to
be older (to have formed the bulk of their stars at earlier times) and
to have formed over shorter time-scales than their lower-mass
counterparts (e.g. \citealt{heavens04, kodama04, juneau05, nelan05,
  thomas05, jimenez07, smith07}).  This has come to be called `archaeological downsizing'.  The results of these studies are
consistent with those which demonstrate that $\sigma$ is the best
predictor of the SFHs of ETGs over $L$, $M_{\star}$, or dynamical mass
$M_{dyn}$ (\citealt{trager00b}; Graves, Faber \& Schiavon 2009a,b; \citealt{rogers10, vanderwel09}; Wake, van Dokkum \& Franx 2012).

The SFHs of ETGs, however, are not purely a 1D family, but are observed to compose at least a 2-parameter family (\citealt{trager00b, graves09b}; Graves, Faber \& Schiavon 2010; \citealt{springob12}).  \citet{graves10b} showed
the stellar populations to map onto the
Fundamental Plane (FP) of ETGs (see e.g. \citealt{dressler87, djorgovski87}; J\o rgensen, Franx \& Kj\ae rgaard 1996) such that it is possible to estimate the SPPs and, thus, the SFHs of quiescent ETGs by their locations in FP-space.  The stellar
populations of quiescent ETGs were shown to scale
systematically with two relevant structural parameters: $\sigma$ and
surface brightness residuals from the FP, $\Delta I_{e}$.  Mean light-weighted age, [Fe/H], [Mg/H], and [Mg/Fe] all increase with increasing $\sigma$, while at
fixed $\sigma$, as $\Delta I_{e}$ increases, [Fe/H] and [Mg/H] increase while age and [Mg/Fe] decrease.  These results led the authors
to propose a premature truncation model in which the onset time and
duration of star formation in quiescent ETGs depend on
$\sigma$ such that higher-$\sigma$ galaxies are older and were formed
over shorter time-scales, while at fixed $\sigma$, galaxies offset to lower
$\Delta I_{e}$ had star formation truncated earlier than those offset
to higher $\Delta I_{e}$.  Thus, according to this model, star
formation is similar for galaxies at similar $\sigma$, while their
SPPs can still vary according to differences in
truncation time, in a way that scales systematically with $\Delta
I_{e}$.  Furthermore, stellar mass-to-light ratios at fixed $\sigma$
are nearly constant \citep{graves10a}.  Thus the $I_{e}$ variations
are primarily variations in stellar mass surface density
$\Sigma_{\star}$, and high-$I_{e}$ galaxies are in fact galaxies with
high $\Sigma_{\star}$ and (typically) high $M_{\star}$ for a given
$\sigma$.

If the SFHs of ETGs are a 2D family, then studies characterizing trends
in the SPPs with environment at fixed $M_{\star}$ vs. at fixed $\sigma$,
for example, will in general yield different and ambiguous results.  We
therefore study the environmental dependence of the SPPs of ETGs at
fixed position in FP-space (along fixed structural parameters shown by
\citet{graves10b} to control the SFHs of ETGs).  This allows us to
identify stellar population differences due solely to environment, as
opposed to those due also to differences in galaxy structure as a
function of environment.

To quantify galaxy environment we use the halo-based group-finding
algorithm of \citet{yang07},\footnote[1]{We use the results of the
  group-finding algorithm of \citet{yang07} applied to SDSS DR7.} which
assigns individual galaxies to their respective dark matter haloes,
assigns halo masses, and distinguishes between the most massive galaxies
in groups, and satellites.  Previous authors have introduced various
other measures of galaxy environment.  Two of the most common of these
are the projected number density of galaxies above a given magnitude
limit (e.g. \citealt{dressler80, lewis02, gomez03, goto03, balogh04a,
  balogh04b, tanaka04, cooper08, cooper10, cooper12}) and the clustering
strength of galaxies using the two-point correlation function
(e.g. \citealt{wake04, croom05, li06, vandenbosch07}) or a marked
correlation function (e.g. \citealt{beisbart00, sheth04, skibba06,
  skibba09}).  The first of these has the disadvantage that its physical
interpretation depends on the environment itself (see
\citet{weinmann06}), while the latter has the disadvantage that it
assigns halo masses to galaxies in a statistical sense, rather than for
individual galaxies \citep{pasquali10}.  We use the halo-based
group-finding algorithm of \citet{yang07} because it is free of these
ambiguities and provides an intuitive measure of an individual galaxy's
environment.

\citet{rogers10} recently used the galaxy group catalogues of
\citet{yang07} to study differences in local ETG SPPs with environment
(central vs. satellite and group halo mass $M_{H}$) at fixed $\sigma$,
and found centrals to have younger ages and significant recent star
formation compared to satellites of the same $\sigma$.
\citet{pasquali10} conducted a similar study for local galaxies at
fixed $M_{\star}$, and found satellite galaxies to be older and more
metal-rich than centrals at fixed $M_{\star}$ (we note, however, that
comparison of the results we present here with those of \citet{pasquali10} is
ambiguous because the authors did not make any morphological or
emission cut to their galaxy sample, whereas we study a sample
of spectroscopically early-type galaxies).  These studies, however, characterized the environmental dependence of the SPPs only along galaxy mass, whereas we have said that the SPPs have been shown to comprise at least a 2D family, with at least two structural controlling parameters ($\sigma$ and $I_{e}$).  We therefore ask a similar question as \citet{rogers10} and \citet{pasquali10} except, extending upon the work of \citet{graves10b},
we study differences in the stellar population properties of quiescent
early-type galaxies with environment at fixed position in FP-space.
When we compare galaxies at the same place in FP-space, do there exist
further trends in these SPPs with environment?  The answer to this
question has important implications for the formation processes of ETGs
in different environments, as manifested in their SFHs.

We select a sample of 28,954 quiescent early-type galaxies from the SDSS
DR7.  We map these galaxies and their derived stellar population
properties age, [Fe/H], and [Mg/Fe] onto and through the Fundamental
Plane.  We also divide our galaxy sample into three classifications of
environment, derived from the galaxy group catalogues of \citet{yang07}.
After confirming the trends seen in the SPPs with the relevant FP
parameters by \citet{graves10b} in our own sample, we then go on to
quantify any differences in the SPPs of our sample with environment at
fixed FP position.

In section 2 we describe the data used in this analysis, including sample selection.  In section 3 we describe our analysis, including sample classification and stellar population analysis.  In section 4 we present our results for variations in the stellar population properties of our quiescent early-type galaxy sample with local group environment at fixed structure.  In section 5 we discuss our results in the context of a few late-time evolutionary processes.  Finally, section 6 summarizes our conclusions.

\section{Data}
The sample of galaxies analysed here are taken from the SDSS Data
Release 7 (DR7) Main Galaxy Survey \citep{abazajian09}.  DR7 marks the
completion of the survey phase known as SDSS-II.

Basic photometric and spectroscopic galaxy properties, including
redshifts and stellar velocity dispersions, are taken from the New York
University Value-Added Galaxy Catalog (NYU-VAGC; \citet{blanton05a}).
Following e.g. \citet{bernardi03}, velocity dispersions are aperture
corrected to 1/8 effective radius using $\sigma_{corr}$ = $\sigma_{fib}
\left( r_{fib}/ \frac{1}{8} r_{0} \right)^{0.04}$, where $r_{0}$ is the
circularized radius in arcseconds and $r_{fib}$ is the spectral fiber
radius (1.5'' for SDSS spectra).  deVaucouleurs radii and axis ratios
are taken from the SDSS DR7 Catalog Archive
Server,\footnote[2]{http://cas.sdss.org/dr7/en/} and are converted to
physical radii (in kpc).  Radii are circularized using $r_{0}$ =
$r_{deV} \sqrt{ \left( b/a \right)_{deV}}$, where $r_{deV}$ and $\left(
b/a \right)_{deV}$ are the deVaucouleurs radius and axis ratio,
respectively.  H$\alpha$, {\oII}$\lambda$3726 and {\oII}$\lambda$3729
emission line fluxes and continua are taken from the MPA-JHU spectral
catalogs.\footnote[3]{http://www.mpa-garching.mpg.de/SDSS/DR7/}  $ugriz$
absolute magnitudes are computed and K-corrected to $z$=0 using the IDL
code $\emph{kcorrect}$ \citep{blanton07}.

Galaxy group information (central vs. satellite, group size, group halo
mass) is taken from the Yang et al. group catalogs.  The SDSS DR7 galaxy
group catalogue is similar to the SDSS DR4 group catalogue of
\citet{yang07}.  The DR7 group catalogue is constructed by applying the
halo-based group finder of \citet{yang05} to the NYU-VAGC.  From the
NYU-VAGC, Yang et al. selected all galaxies in the Main Galaxy Sample
with an extinction-corrected apparent magnitude brighter than $r$ =
17.72, with redshifts in the range 0.01 $\leq$ $z$ $\leq$ 0.20 and with
a redshift completeness $C_{z}$ $>$ 0.7.  Of the three group samples
Yang et al. construct from this parent sample, we use Sample II.  In
addition to the 599,301 galaxies in Sample I with measured $r$-band
magnitudes and redshifts from SDSS, Sample II includes 3,269 galaxies
with spectroscopic redshifts taken from alternative surveys.

Our sample contains galaxies restricted to a redshift range 0.025 $<$ $z$ $<$ 0.1 and a $\sigma$ range 70 (km/s) $<$ $z$ $<$ 300 (km/s).  Where necessary we assume a standard $\Lambda$CDM cosmology with $H_{0}$=73 km/s/Mpc, $\Omega_{M}$=0.25, and $\Omega_{\Lambda}$=0.75.

\subsection{Quiescent Selection}
In the present analysis we study the star formation histories of
quiescent early-type galaxies.  We construct our quiescent, early-type
galaxy sample by making cuts in the spectral emission features and
colors of our parent galaxy sample.

Our emission line cuts are based on H$\alpha$ and {\oII}$\lambda$3727,
which are typically the two strongest optical emission lines in
star-forming galaxies.  This cut also eliminates contamination from
low-ionization nuclear emission-line regions (LINERs), which exhibit
high {\oII}/H$\alpha$ ratios, and galaxies which exhibit low
{\oII}/H$\alpha$ ratios, including active galactic nuclei (AGN),
transition objects (TOs), and Seyferts \citep{yan06}.  These cuts
largely parallel those made in \citet{peek10}.

\begin{figure*}
\includegraphics[scale=0.7]{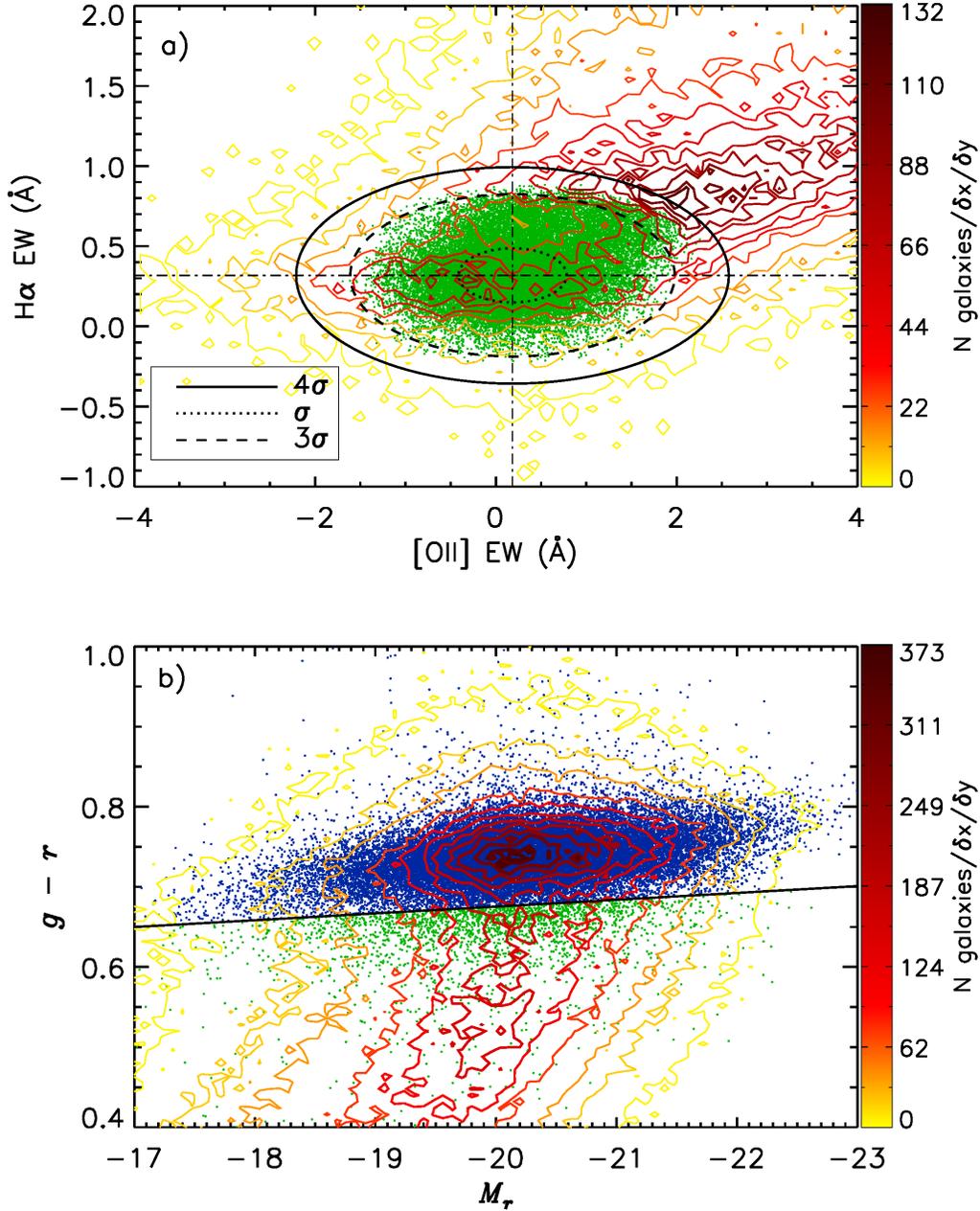}
\caption{\textbf{(a)} The distribution of our parent galaxy sample around zero emission in H$\alpha$ and {\oII}. \textbf{Dashed-dotted lines} indicate the zero-point centre of EW measurements (0.184,0.318).  \textbf{Ellipses} centred at the zero-point of the quiescent distribution whose semimajor and semiminor axes are 1, 3, and 4 times the Gaussian widths of the quiescent distribution (0.598 $\AA$, 0.168 $\AA$) are indicated. Galaxies which are selected by our emission cuts (\textbf{green}) are those whose measured emission values $\pm$1$\sigma$ lie within the ellipse with semimajor and semiminor axes equal to 4 times the Gaussian widths of the quiescent distribution (\textbf{solid line}). The smoothed distribution of galaxies eliminated from our sample by the EW selection is shown with \textbf{contour lines}, plotted in different colors to show the density of the galaxy distribution in this space according to the scale shown in the \textbf{colorbar} to the right of the panel.  Bin sizes along the horizontal and vertical dimensions used to smooth the eliminated galaxy population are $\delta \left( \text{{\oII} EW} \right)$=0.13 $\AA$ and $\delta \left( \text{H}\alpha \text{ EW} \right)$=0.06 $\AA$. \textbf{(b)} The distribution of our parent galaxy sample in color-magnitude space. Galaxies that lie below the \textbf{solid line} parallel to the red sequence, offset below it by 2$\sigma_{y,rot}$=0.021, where 2$\sigma_{y,rot}$ is the Gaussian width of the red sequence distribution along the direction perpendicular to the red sequence, are eliminated by our color cuts.  The smoothed distribution of galaxies eliminated by our emission cuts is shown with \textbf{contour lines}, now showing the density of galaxies in color-magnitude space according to the scale shown in the \textbf{colorbar} to the right of the panel.  Bin sizes along the horizontal and vertical dimensions used to smooth the galaxy population in this space are $\delta \left( M_{r} \right)$=0.08 and $\delta \left( g - r \right)$=0.008.  Galaxies which pass our emission cuts but not our color cuts are shown in \textbf{green}, and the 28,954 galaxies which pass both cuts and define our final sample are shown in \textbf{blue}.} \label{cuts}
\end{figure*}

Figure \ref{cuts}$\mathrm{a}$ shows the distribution of the full $0.025 < z < 0.1$ galaxy sample in H$\alpha$ and {\oII} equivalent width (EW), zoomed in
around the region of galaxies with zero emission.  The quiescent
galaxies used in the present analysis are those that lie near the peak
of the distribution around zero emission in both H$\alpha$ and {\oII}.
The population of truly quiescent galaxies should be symmetrically
distributed around (0,0), with scatter due to measurement errors in the
emission line strengths.  There may be zero-point offsets in the EW
measurements (see, e.g., \citealt{yan06}), and the typical errors in
H$\alpha$ EW are different from those in {\oII}.  With these effects in
mind, we determine the EW zero-point values and selection region as
follows.

We smooth over the distribution of galaxies in the H$\alpha$-{\oII}
EW-space by taking its two-dimensional histogram.  We isolate quiescent
galaxies in this space by fitting a Gaussian with zero tilt around the
distribution peak.  This Gaussian fit is centred at (0.184,0.318),
which we take to indicate the zero-point of the EW measurements.  The
Gaussian widths are 0.598 $\AA$ and 0.168 $\AA$ in {\oII} and H$\alpha$
EW, respectively.  We then make a cut in emission to the galaxy
population in EW-space: selected galaxies are those whose measured
values $\pm$1$\sigma$ lie within the ellipse centred at (0.184,0.318)
with width and height equal to 4 times the Gaussian widths of the
quiescent distribution.

These selection criteria minimize contamination from galaxies with
significant H$\alpha$ and {\oII} emission, while still maintaining a
large enough sample for later analysis.  The galaxies which meet these
criteria are shown in green in Figure \ref{cuts}$\mathrm{a}$, while the
distribution of those which are eliminated from our sample by the EW
selection is shown with contour lines.  Figure \ref{cuts}$\mathrm{a}$ also shows ellipses centred at
the Gaussian centre of the quiescent distribution whose semimajor and
semiminor axes are 1,3, and 4 times the Gaussian widths of the quiescent
distribution.  Making our cut is roughly equivalent to making a cut such
that galaxies lie within the 3$\sigma$ ellipse with no consideration of
their errors in emission.

After this emission cut we make a further cut in color-magnitude space
to remove color outliers.  This identifies contaminants to our passive
sample, e.g., galaxies where the emission line measurements have failed.
We first define the red sequence of our full galaxy sample in $g-r$
vs. $M_{r}$ color-magnitude space (Figure \ref{cuts}$\mathrm{b}$) in the
following way.

We again take a two dimensional histogram of the galaxy distribution,
this time in ($g-r$) vs. $M_{r}$ space.  We then fit a two-dimensional
Gaussian function with nonzero tilt to the galaxy distribution centred
around the red sequence.  We define the centre of the red sequence in
color-magnitude space to be along a line connecting the major axes of
ellipses that are the level sets of our 2D-Gaussian fit to the red
sequence.  We then define
a line parallel to this red sequence, offset below it by
2$\sigma_{y,rot}$ (solid line in Figure \ref{cuts}$\mathrm{b}$), where $\sigma_{y,rot}$=0.021 is the Gaussian width of
the red-sequence distribution along the direction perpendicular to the red sequence line.  Galaxies that lie below this solid line
are discarded.

Our cut is illustrated in Figure \ref{cuts}$\mathrm{b}$, where our
parent sample of galaxies is shown in color-magnitude space separated
into three subpopulations: the distribution of galaxies which were
eliminated by our emission cuts is shown with contour lines; galaxies which passed the emission cuts but not the color cuts are shown in green; galaxies which passed both cuts and define our final quiescent red-sequence sample are shown in blue.  A visual inspection of the spectra for objects plotted in green (those with no emission but blue colors) showed that many of them contain significant emission and thus represent errors in the MPA-JHU emission line measurements or mismatches between the catalogs.  Our color cuts yield a sample that is not significantly contaminated by objects lying off the red sequence.  After all the cuts described above we are left with a sample of 28,954 galaxies.

\section{Analysis}

\subsection{Defining 3D Fundamental Plane Space}
In the present analysis we study how the stellar
population properties and, thus, the star formation histories of
quiescent early-type galaxies relate to their present-day structural
properties and local group environments.  In order to understand how
these SFHs scale with present-day structural properties we study their
variation along and `through' the thickness of the Fundamental Plane.

Here we describe our 3D FP binning strategy.  We follow the method of
\citet{graves09b} by dividing galaxies into this space directly using
the FP parameters $\log R_{e}$, $\log \sigma$, and $\log I_{e}$.
Although these parameters are not orthogonal, we use them as bases and
bin galaxies in this space because we can more directly interpret the
structural properties of galaxies in our sample by their locations in
this space.

\begin{figure*}
\includegraphics[scale=1.0]{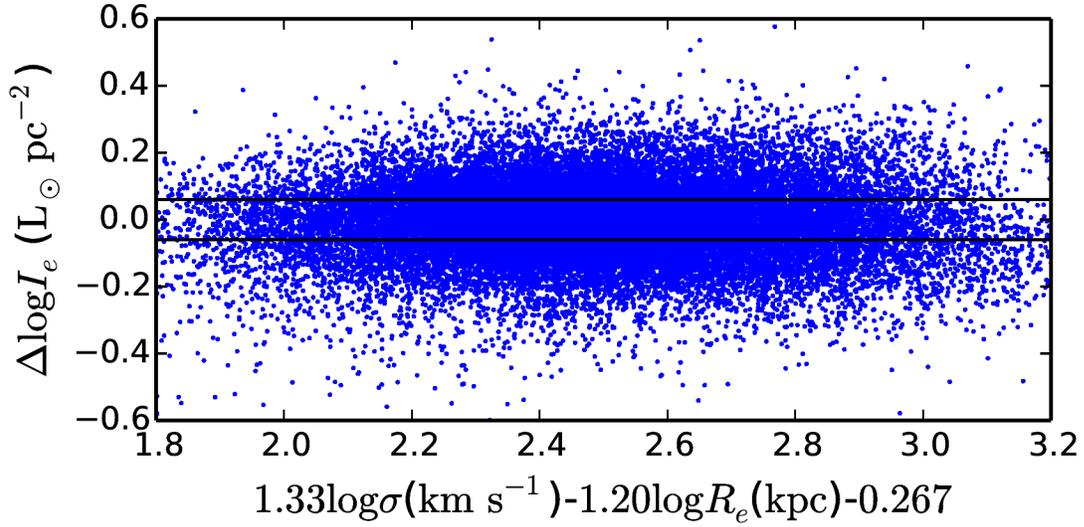}
\caption{An edge-on view of the Fundamental Plane (FP) in the $I_{e}$ direction. Scatter off of and perpendicular to the FP constitutes the $\Delta \log I_{e}$ dimension (thickness) of our FP space.  The measured value of $\Delta \log I_{e}$ of each galaxy in our sample is shown plotted against its value of $\log I_{e}$ predicted from our plane fit to the $\log I_{e}$-$\log R_{e}$-$\log \sigma$ relation.  \textbf{Solid lines} indicate our division of the thickness of the FP into `Low-SB' ($\Delta \log I_{e}$ $\leq$ -0.059), `Mid-SB' (-0.059 $<$ $\Delta \log I_{e}$ $<$ 0.059), and `High-SB' ($\Delta \log I_{e}$ $\geq$ 0.059) slices.} \label{edge}
\end{figure*}

We construct the $\Delta I_{e}$ dimension (thickness) of our FP-space by fitting a plane to the $\log I_{e}$-$\log R_{e}$-$\log \sigma$ relation.  This fit uses the routine $\emph{leastsq}$ in the package $\emph{scipy.optimize}$ in Python to minimize residuals in the $\log I_{e}$ direction.  We find the following FP relation from this fit:
\begin{equation}
\log I_{e} = 1.33\log \sigma - 1.20\log R_{e} - 0.267,
\label{FP_eq}
\end{equation}
where $I_{e}$ is measured in $L_{\odot}$ $\mathrm{pc}^{-2}$, $\sigma$ is measured
in km s$^{-1}$, and $R_{e}$ is measured in kpc.  The FP residual,
$\Delta \log I_{e}$, is the difference between the measured $\log
I_{e}$ and that expected from the FP fit, i.e., $\Delta \log I_{e}$
$\equiv$ $\log I_{e}$ - ($1.33 \log \sigma$ - $1.20 \log R_{e}$ -
$0.267$).  An edge-on view of the FP is shown in Figure \ref{edge}.

As demonstrated in \citet{graves10a}, the thickness in $\Delta I_{e}$
of the FP is predominantly the result of variations in the stellar
surface mass density $\Sigma_{\star}$.  This can be interpreted as a
conversion efficiency for turning gas into stars.  In order to study
variations in the SPPs with $\Delta I_{e}$, we define three layers
through the FP by dividing $\Delta \log I_{e}$ into three `slices':
`Low-SB' ($\Delta \log I_{e}$ $\leq$ -0.059), `Mid-SB' (-0.059 $<$
$\Delta \log I_{e}$ $<$ 0.059), and `High-SB' ($\Delta \log I_{e}$
$\geq$ 0.059).  The three layers along the $\log I_{e}$ dimension,
separating galaxies in this space into Low-, Mid-, and High-SB are
indicated by black lines in Figure \ref{edge}.

In each of the three surface brightness slices, we then bin galaxies
into a 6$\times$5 grid in the $R_{e}$-$\sigma$ projection of the FP
(see any panel of Figures \ref{age} through \ref{mg_fe}).
This yields 30 bins of galaxies in the $R_{e}$-$\sigma$ space, and a
total of 90 bins of galaxies in our 3D FP-space defined by the three
parameters $I_{e}$, $R_{e}$, and $\sigma$.  This $R_{e}$-$\sigma$
projection of FP-space also conveys important information about the
structures of our galaxy sample.  By fixing the position of a galaxy in
this projection of FP-space we fix its total mass ($M_{dyn}$ $\propto$
$\sigma^{2} R_{e}$), and by varying the position of a galaxy along a
line of constant $M_{dyn}$ we vary its size and concentration.

This FP-space provides a description of a galaxy's structure along which we can study changes in the SFHs of our galaxy sample in different environments.

\subsection{Classifying Local Group Environment}
This analysis aims to study the role of local group environment in the
star formation histories of early-type galaxies.  Here we use `local
group environment' to refer to a galaxy's `identity' within its dark
matter halo, i.e., whether it is the central most massive galaxy within
a halo hosting other less massive galaxies, a smaller satellite galaxy
within the halo of a more massive host galaxy, or the only galaxy in its
dark matter halo.  This provides an intuitive measure of an individual
galaxy's environment.  We ask the question whether, when we fix
quiescent ETGs by their locations in FP-space (along the structural
parameters shown by \citet{graves10b} to be well-correlated to their
SFHs), their SPPs are seen to vary with their local group environments.

To describe local group environment we classify galaxies into three
categories: `Brightest Group Galaxies' (BGGs) are defined as the single
most massive galaxies residing in groups containing more than one
galaxy;\footnote[4]{Note that the classification of BGG is based on
  $M_{\star}$, not on $L$.  These galaxies are therefore truly the
  `most massive group galaxy'.  We use the term BGG in analogy with the
  more familiar `BCG' for `Brightest Cluster Galaxy'.} Satellites are
defined as those galaxies that are not the most massive in a group
containing more than one galaxy (they reside in the dark matter halo of
the host galaxy); Isolateds are those galaxies that are the only members
of the dark matter haloes in which they reside.\footnote[5]{`Isolated'
  galaxies can in theory have satellites that fall below the sample
  magnitude limit.}  By defining these three group classifications we
further divide our 90 bins in FP-space into a total of 270 bins.

To make these classifications we use the DR7 version of the group catalogue of \citet{yang07}.  The Yang group catalogue is constructed using a halo-based group finder, which establishes groups by assigning individual galaxies to their respective dark matter haloes.  The halo-based group finder does this by first making tentative group assignments using the traditional FOF algorithm of \citet{davis85}, then assigning a tentative mass to each group with an assumption of the group mass-to-light ratio.  This mass is then used to estimate the size and velocity dispersion of the host halo, which is then used to redetermine group membership in redshift space.  This process is iterated until group membership converges.  For a full description of the group finding algorithm and the basic properties of the group catalogue please see the description of the earlier DR4 group catalogue \citep{yang07}.  We use the DR7 group catalogue rather than the DR4 group catalogue because it gives us a larger sample size.

\subsection{Constructing the Stacked Spectra}
Our 3D FP-space, divided into BGGs, Isolateds, and Satellites, contains 270 bins of galaxies.  For each of these 270 bins, we co-add the spectra of the individual
galaxies within the bin to make a very high S/N stacked spectrum.
The individual spectra are normalized in the 4000--5500{\AA}
wavelength region and combined in an unweighted sum, so that each
constituent galaxy contributes equally to the final stack.  We mask
out regions around strong skylines, and perform pixel-by-pixel
$\sigma$-clipping to remove distant outliers. We smooth all the
spectra up to a common effective velocity dispersion to match the
resolution of the highest-$\sigma$ galaxy within each bin.  Corresponding
error spectra are constructed in parallel.  The total S/N varies,
depending on the number of galaxies that contribute to each bin. We
note that prior to stacking, galaxy spectra are not corrected for
Galactic foreground extinction, which is expected to have a negligible
effect.\footnote[6]{http://www.sdss.org/dr7/products/spectra/index.html} After our cuts in emission and in color-magnitude space, we are still
left with a small number of galaxies that contain significant
emission, whose catalog emission values do not correctly reflect the
properties of the spectra.  These are a small fraction of the total
galaxies.  Thus in bins containing many galaxies, the emission lines
are effectively excluded by the $\sigma$-clipping algorithm.  However,
in bins with relatively few galaxies, no pixels are removed by
$\sigma$-clipping and the mis-identified objects can contribute
substantial emission to the stack.  To eliminate these objects, we
make iterative manual checks to the stacked stellar spectra for each
bin.  If any substantial emission is seen in the stacked spectrum, we
visually inspect the spectrum of each individual galaxy belonging to
that bin, eliminating from our sample those galaxies whose spectra
show significant emission that was not reflected in their catalog
emission line values.  These iterative checks remove 36 galaxies from
our sample of $\sim$29,000 galaxies.

\subsection{Stellar Population Analysis}
The stacked spectra for each bin can be used to characterize the
stellar population properties that are typical for galaxies in that
bin.  We use the IDL code EZ\_Ages \citep{graves08}, based on the
stellar population synthesis models of \citet{schiavon07}, to measure
single burst ages, [Fe/H], and [Mg/Fe] for each stacked spectrum.
Briefly, EZ\_Ages uses Balmer and Fe absorption lines to fix a
fiducial age and [Fe/H] for the spectrum, based on Solar-abundance
model grids.  It then adjusts the abundance of [Mg/Fe] with respect to
this Solar-scale model until the model can consistently fit the Mg
{\it b} absorption feature as well.  The code then iterates to ensure
a self-consistent solution.  Other absorption lines can be used to fit
other elements, but here we restrict ourselves to fitting age, [Fe/H],
and [Mg/Fe].  EZ\_Ages also provides error estimates for these
parameters, based on the formalism of \citet{cardiel98}, which has
been shown to be consistent with the error estimates from Monte Carlo
simulations \citep{graves08}.  For more details on the modeling
process, we refer the reader to \citet{graves08}.
The resulting stellar population measurements give a light-weighted single-burst age,
mean [Fe/H], and [Mg/Fe].  Each of these parameters contains
information about the star formation histories of the galaxies in
question.

The measured stellar population ages are derived from model star
formation histories that consist of a single burst of star formation.
Single burst models are not realistic representations of galaxy star
formation histories, and certainly not for a stack of many galaxies.
Thus the single burst ages quoted here are used as statistical
descriptions of the ensemble of stars in a galaxy bin in order to make
quantitative comparisons between bins, without assuming that the
measured ages correspond to a rigorous physical formation time.  The
reader should bear in mind that the stars in any given galaxy bin
likely formed over a range of times.  Furthermore, the single-burst
ages are not equivalent to mean mass-weighted ages but instead are
strongly skewed by young sub-populations \citep{trager09}.  They thus
capture information about whether or not a galaxy has formed a
substantial quantity of stars in the recent past.
We use the Fe abundance [Fe/H] to quantify the overall normalization
of the metal abundance pattern in our stacked galaxies.  Again, these
are mean quantities over an ensemble of metallicities, although these
are much more nearly mass-weighted quantities \citep{trager09}.
Because the nucleosynthesis of Fe is dominated by Type Ia supernovae
(SNe Ia), [Fe/H] measures the quantity of SNe Ia product that has
built up during a galaxy's star forming lifetime through previous
generations of stars.  
Finally, we use [Mg/Fe] to trace the typical duration of star
formation in our galaxies.  The nucleosynthesis of Mg is dominated by
massive stars and Type II supernovae (SNe II), which have shorter
lifetimes than the stellar progenitors of SNe Ia.  Thus an excess of
Mg with respect to Fe is often interpreted to mean that star formation
in a galaxy was of short duration, such that the interstellar medium
(ISM) had time to enrich in SNe II product (Mg), but not to fully
enrich in SNe Ia product (Fe) (see, e.g., \citealt{tinsley79,
  greggio83}; Worthey, Faber \& Gonz\'{a}lez 1992;
\citealt{matteucci94, trager00a, thomas05, delarosa11}).

\section{Results}
The results of our analysis are shown in Figures \ref{age} through
\ref{mg_fe}.  Here we show the final galaxy sample in 3D FP-space,
separated into BGGs, Isolateds, and Satellites.  Each panel shown in
Figures \ref{age} through \ref{mg_fe} shows the $R_{e}$-$\sigma$
projection of FP-space.  In each of these figures the top, middle, and
bottom rows of panels show galaxies in our High-, Mid-, and Low-SB FP
slices, respectively.  The left, middle, and right columns show our
Isolated, BGG, and Satellite galaxy samples, respectively.  The SPPs
age, [Fe/H], and [Mg/Fe] for each bin in this space are shown by the
color in which a galaxy is plotted, according to the scale shown by the
colorbar in each figure.  In the top, leftmost panels of Figures
\ref{age} through \ref{mg_fe}, a dashed red line of constant dynamical
mass $M_{dyn}$ is shown.

Looking first at the distribution of galaxies in the binned space, we notice that BGGs are distributed differently from the Satellite and Isolated populations.  In a given surface brightness slice BGGs tend to populate the high-$\sigma$, high-$R_{e}$ part of FP-space, where we expect to find the most massive galaxies (recall that $M_{dyn}$ $\propto$ $\sigma^{2} R_{e}$ for a Virialized system).  It is no surprise that BGGs tend to be more massive than Satellites; this is built into their definition.  That BGGs also tend to be more massive than Isolateds is not directly definitional, and illustrates that more massive haloes tend to host more massive central galaxies.

\subsection{Star Formation History vs. Structure and Environment}
\begin{figure*}
\includegraphics[scale=0.65]{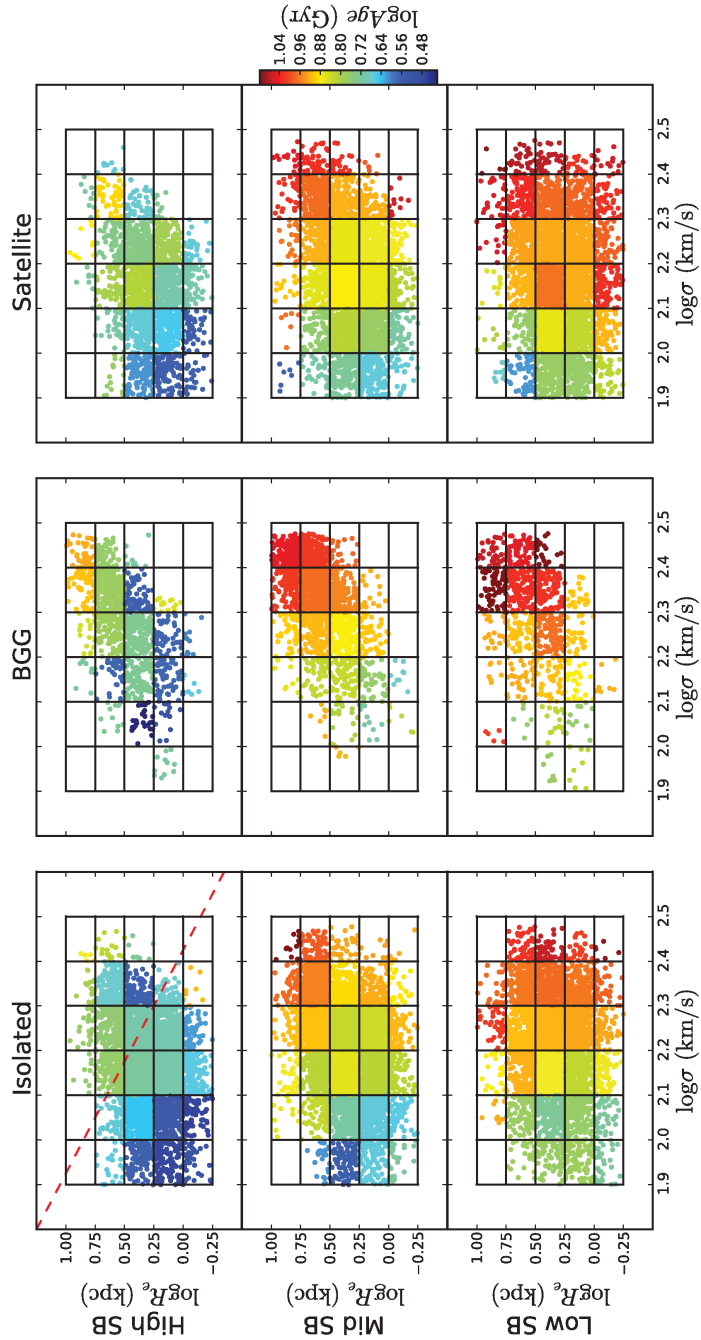}
\caption{Our sample shown in 3D FP-space, separated into
    Isolateds (\textbf{left columns}), BGGs (\textbf{middle columns}),
    and Satellites (\textbf{right columns}).  \textbf{Top},
    \textbf{middle}, and \textbf{bottom rows} show High-, Mid-, and
    Low-SB FP slices, respectively.  Our bins in the $\log R_{e}$-$\log
    \sigma$ FP projection are shown separated by \textbf{solid vertical}
    and \textbf{horizontal lines}.  BGGs tend to occupy the most massive
    end of the space.  Lines of constant $M_{dyn}$ in the $\log R_{e}$-$\log \sigma$ projection follow that shown by a
    \textbf{dashed red line} in the top, leftmost panel ($M_{dyn}
    \propto \sigma^{2} R_{e}$).  Mean stellar population age is shown by
    the color in which each galaxy is plotted, according to the scale
    shown by the \textbf{colorbar} to the right.  To zeroth-order, the
    ages of our sample are fully captured by their structures,
    increasing with increasing $\sigma$, and at fixed $\sigma$,
    increasing with decreasing $\Delta I_{e}$.  Trends in age with
    environment at fixed structure are only slight: Satellites are
    slightly older than BGGs, and both are older than
    Isolateds.} \label{age}
\end{figure*}

\begin{figure*}
\begin{center}
\includegraphics[scale=0.9]{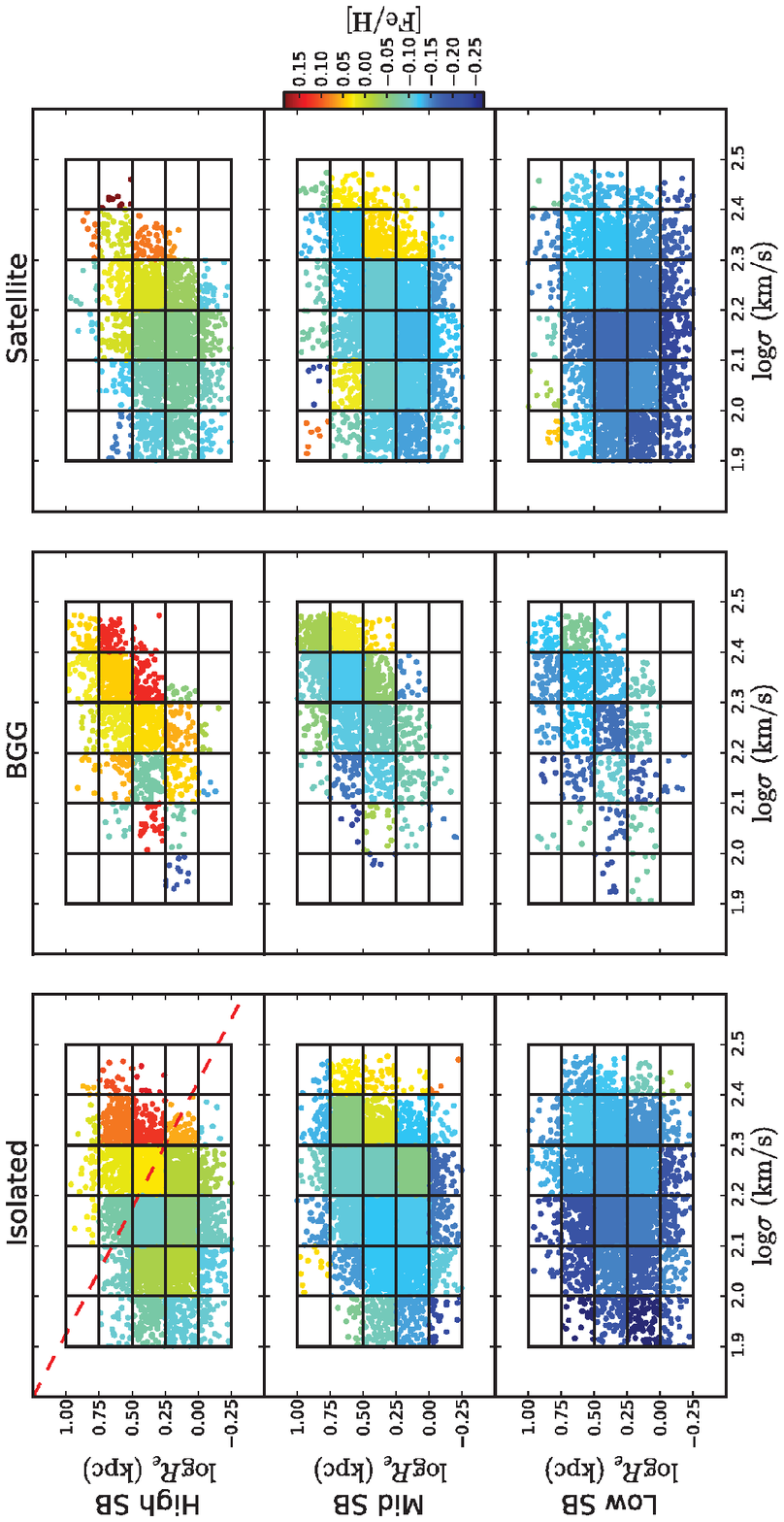}
\caption{Same as Figure \ref{age}, now showing mean
    stellar population [Fe/H] by the color in which each galaxy is
    plotted in FP-space, separated into BGGs, Isolateds, and Satellites.
    To zeroth-order, the [Fe/H] of our sample are fully captured by
    their structures, increasing with increasing $\sigma$, and at fixed
    $\sigma$, increasing with increasing $\Delta I_{e}$.  Trends in
    [Fe/H] with environment at fixed structure are even more slight than
    those found in the ages: BGGs are slightly more Fe-rich than both
    Satellites and Isolateds.} \label{fe_h}
\end{center}
\end{figure*}

\begin{figure*}
\begin{center}
\includegraphics[scale=0.9]{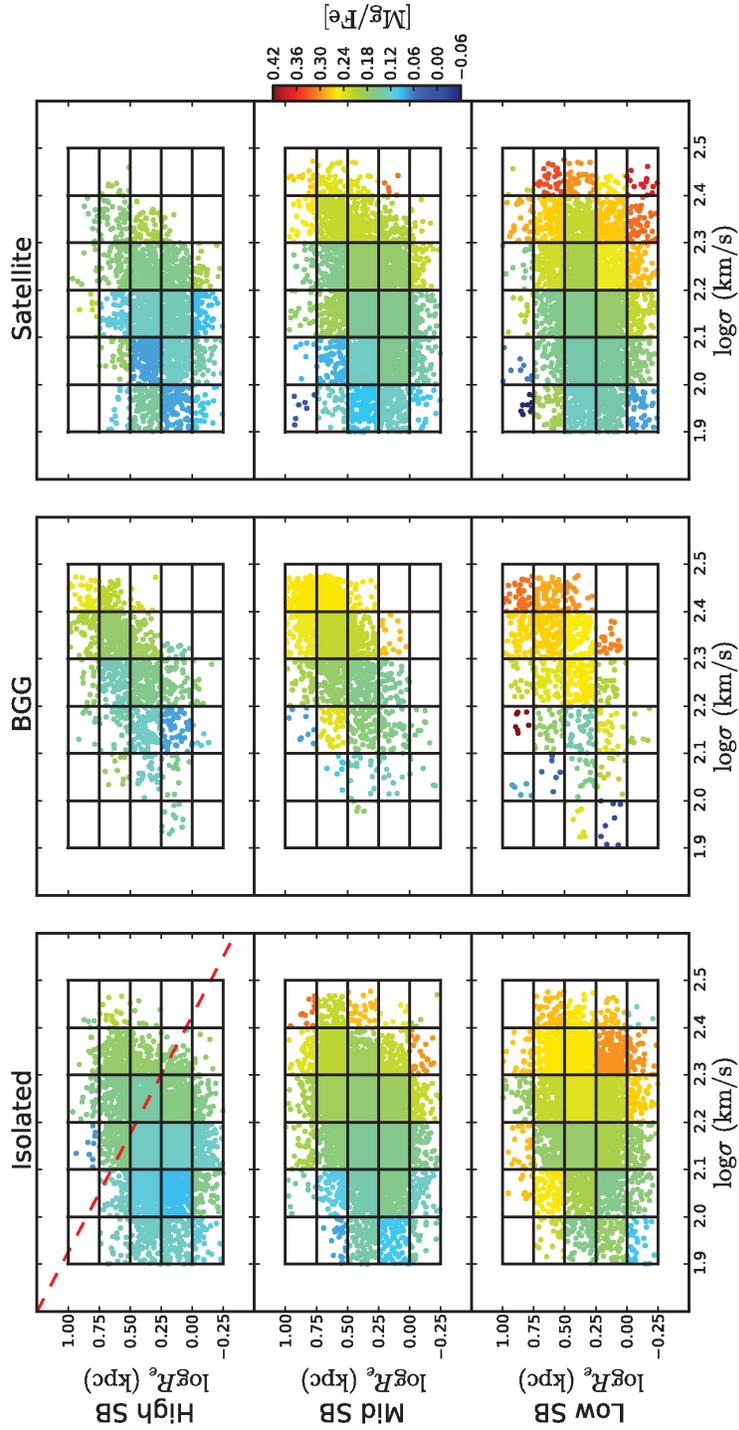}
\caption{Same as Figure \ref{age}, now showing mean
    stellar population [Mg/Fe] by the color in which each galaxy is
    plotted in FP-space, separated into BGGs, Isolateds, and Satellites.
    The [Mg/Fe] of our sample are fully captured by their structures,
    increasing with increasing $\sigma$, and at fixed $\sigma$,
    increasing with decreasing $\Delta I_{e}$.  There are no overall
    trends in [Mg/Fe] with local group environment at fixed
    structure.} \label{mg_fe}
\end{center}
\end{figure*}

We confirm all of the previous trends observed in the
SPPs of quiescent ETGs with
their positions in FP-space (see \citealt{graves10b, springob12}).  Studying
Figures \ref{age} through \ref{mg_fe}, the following trends are unanimously
true for BGGs, Isolateds, and Satellites.  Along the direction of
increasing $\sigma$, the ages, [Fe/H], and [Mg/Fe] of galaxies in our sample are observed to increase.
Along the direction of decreasing $\Delta I_{e}$, the ages and [Mg/Fe] of galaxies in our sample are observed to
increase, while their [Fe/H] are observed to decrease.
None of the SPPs explored here (age, [Fe/H], or [Mg/Fe]) are observed to vary
substantially with $R_{e}$.  In agreement with \citet{graves10b}, the stellar population properties of quiescent early-type galaxies
are shown to be well mapped to their structures, with the most relevant
structural parameters being $\Delta I_{e}$ and $\sigma$.

We
now ask the question whether, on top of the known correlations with structural parameters, the SPPs are
seen to vary with local group environment.  Are the processes which
cause the observed correlations environment-dependent?

When comparing the SPPs with environment in Figures \ref{age} through \ref{mg_fe}, it is apparent that any residual trends with environment are subtle. To
zeroth-order, there are no systematic differences in the stellar
population properties with environment at fixed structure. This is our
central result.

This result is surprising.  Because we expect the efficiency of certain
formation processes, such as cooling flows, to depend strongly on
environment, we might expect the SFHs of ETGs in our sample to vary with
local group environment.  Our observations therefore imply that
formation processes which depend on environment cannot alter the
SFHs/structures of galaxies in our sample so significantly as to produce
significant variations in the SFH-structure correlation we observe with
environment.

On top of our zeroth-order conclusion,
however, we do see some very subtle overall differences in the SPPs with environment: considering Figure \ref{age}, and comparing fixed bins in FP-space
between Isolated, BGG, and Satellite galaxies, it appears that Satellite galaxies are slightly older than BGGs, while both are
older than Isolateds.  This difference is taken over all the bins, and
there remain single outliers to this trend, as we see, for example,
comparing the highest $R_{e}$, highest $\sigma$, Mid-SB bin of
Isolated galaxies with the same bin for BGG and Satellite galaxies.  Looking at Figure \ref{fe_h}, there seems to be an overall trend such that BGGs
are more Fe-rich at fixed position in FP-space than both Satellites
and Isolateds.  This trend is most clearly seen in the High-SB FP slice.
It also seems when considering the High-SB slice in Figure \ref{fe_h} that
Isolateds are more Fe-rich than Satellites.  However, this trend does
not persist through the rest of the FP-space, and in particular, in the
Mid-SB slice the trend seems to be the opposite -- Satellites appear more
Fe-rich than Isolateds.  Considering Figure \ref{mg_fe}, it is difficult
to detect any clear trend in the Mg-enhancements of our galaxy sample
with local group environment at fixed position in FP-space.  Satellite
galaxies at the high-$\sigma$, Low-SB bins of FP-space seem to be more
Mg-enhanced than both Isolateds and BGGs, but this trend does not
persist through the rest of the space and is contradicted at various
other positions in FP-space.  We note that uncertainties in the SPPs are not shown in Figures \ref{age} through \ref{mg_fe},
so we cannot make any quantitative judgements regarding subtle
differences in the SPPs with environment here.  This is done in the next
section.

We repeat our main result: that to
zeroth-order, the star formation histories of quiescent early-type
galaxies are captured by their structures, with the two most relevant
structural parameters being $\Delta I_{e}$ and $\sigma$, in the way
described by \citet{graves10b}.  At fixed structure, any differences
in the star formation histories of early-type galaxies with
environment are only slight.  This has interesting implications for
those galaxy formation processes that are expected to depend strongly
on environment. Our observations constrain the degree to which such
processes can alter the SFHs/structures of early-type galaxies in our
sample.

\subsection{Quantifying Residual Trends with Local Group Environment}
\begin{figure*}
\begin{center}
\includegraphics[scale=0.6]{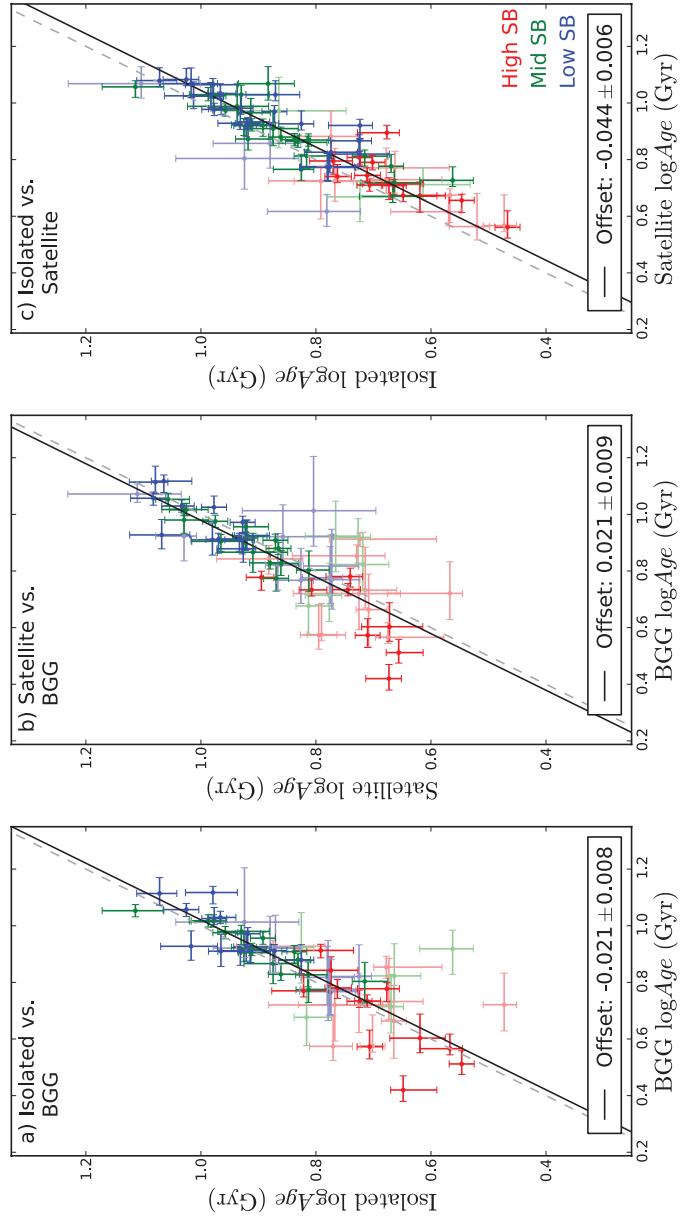}
\caption{The mean ages of each bin in FP-space.  The ages
    of Isolateds vs. BGGs, Satellites vs. BGGs, and Isolateds
    vs. Satellites are shown in panels \textbf{(a)} through
    \textbf{(c)}, respectively.  Points plotted in \textbf{red},
    \textbf{green}, and \textbf{blue} represent High-, Mid-, and Low-SB
    bins, respectively.  Bins that are High-SN in age (whose errors in
    age are less than or equal to 20$\%$ their age values) are shown in
    \textbf{dark shades} of red, green, and blue, while Low-SN age bins
    are shown in \textbf{light shades}.  Our linear fit to High-SN bins
    with slope fixed to 1 is shown with a \textbf{solid line} in each
    panel.  The corresponding one-to-one relation with no offset is
    shown with a \textbf{dashed grey line} for comparison.  The offsets
    we obtain between the ages of each galaxy population are shown at the bottom of each panel.  There are only modest shifts
    in age at fixed structure such that age$_{Sat}$ $>$ age$_{BGG}$ $>$
    Age$_{Iso}$.} \label{age_pp}
\end{center}
\end{figure*}

\begin{figure*}
\begin{center}
\includegraphics[scale=0.65]{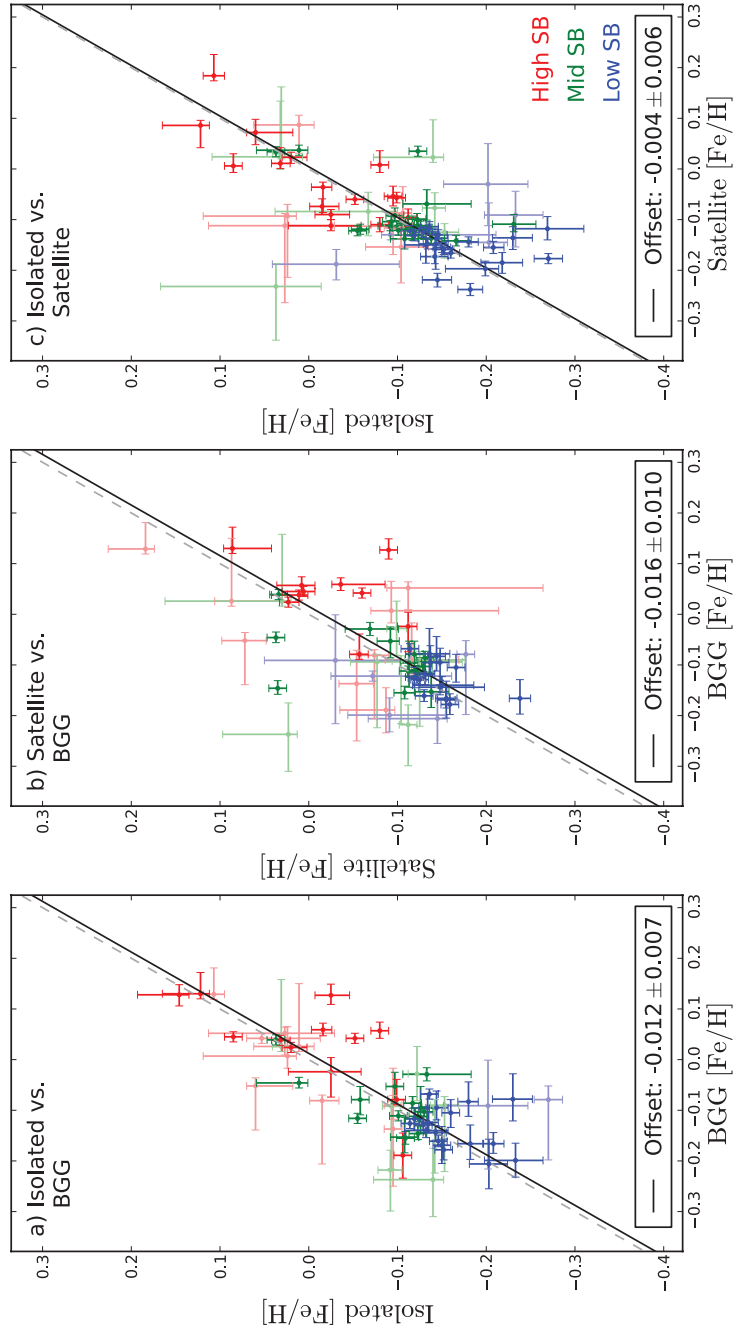}
\caption{Same as Figure \ref{age_pp}, now showing the mean
    [Fe/H] of each bin in FP-space.  Bins that are High-SN in [Fe/H]
    (whose errors in [Fe/H] are less than 0.05 dex) are shown in
    \textbf{dark shades} of red, green, and blue, while Low-SN [Fe/H]
    bins are shown in \textbf{light shades}.  There are very slight
    shifts (\textbf{solid line}) in [Fe/H] at fixed structure such that
    [Fe/H]$_{BGG}$ $>$ [Fe/H]$_{Sat}$ = [Fe/H]$_{Iso}$.} \label{fe_h_pp}
\end{center}
\end{figure*}

\begin{figure*}
\begin{center}
\includegraphics[scale=0.65]{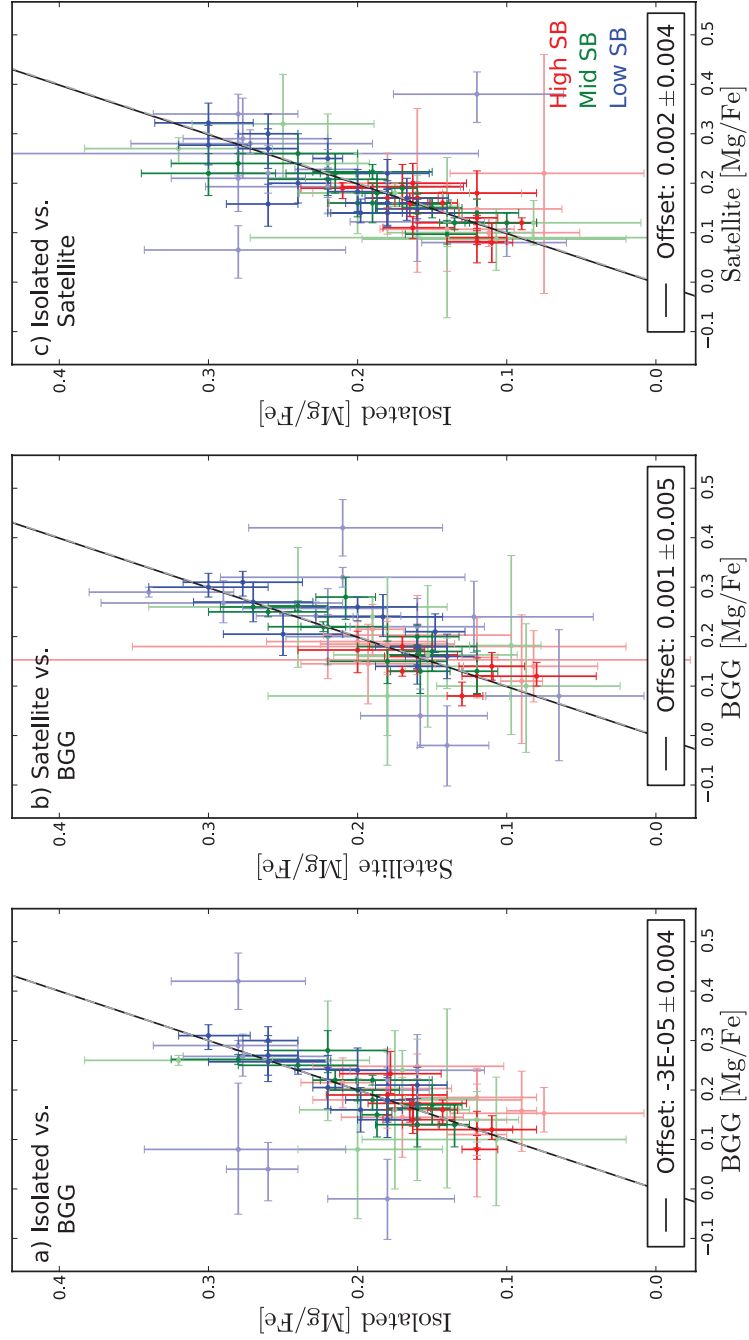}
\caption{Same as Figure \ref{age_pp}, now showing the mean
    [Mg/Fe] of each bin in FP-space.  Bins that are High-SN in [Mg/Fe]
    (whose errors in [Mg/Fe] are less than 0.05 dex) are shown in
    \textbf{dark shades} of red, green, and blue, while Low-SN [Mg/Fe]
    bins are shown in \textbf{light shades}.  The [Mg/Fe] are equal
    between BGG, Isolated, and Satellite galaxies (\textbf{solid
      line}).} \label{mg_fe_pp}
\end{center}
\end{figure*}

To better quantify the behavior of the SPPs of our quiescent ETG sample with local group environment at fixed position in FP-space, we
systematically compare BGG, Isolated, and Satellite galaxy age, [Fe/H], and [Mg/Fe] at each fixed bin of
our FP-space.  The results are shown in Figures \ref{age_pp} through \ref{mg_fe_pp}.  Each point in the plots
shown in Figures \ref{age_pp} through \ref{mg_fe_pp} represents an individual
bin in FP-space.  Points plotted in red, green, and blue represent High-, Mid-, and Low-SB bins, respectively.  We also distinguish between high-SN and low-SN
bins, defined according to their uncertainties for each SPP, in Figures \ref{age_pp} through \ref{mg_fe_pp}.  High-SN
data are shown in dark shades of red, green, and blue while low-SN data
are shown in light shades to reduce their visual impact.  High-SN age bins
are those whose errors in age are less than or equal to 20$\%$ their
age values, while High-SN [Fe/H] and High-SN [Mg/Fe] bins are those whose errors in [Fe/H]
and [Mg/Fe] are less than or equal to 0.05 dex each.
This identifies 69$\%$, 71$\%$, and 57$\%$ of our bins as being High-SN
in age, [Fe/H], and [Mg/Fe],
respectively.

In order to find shifts between the SPPs of galaxies in different environments, we make linear fits to
the High-SN data shown in each of the panels of Figures \ref{age_pp} through \ref{mg_fe_pp}, fitting for the offset $a$ of the line with a slope fixed to
1 by minimizing the $\chi^{2}$ merit function:
\begin{equation}
\chi^{2} = \displaystyle\sum_{i=1}^{N} \frac{\left( y_{i} - a - x_{i} \right)^{2}}{\sigma_{yi}^{2} + \sigma_{xi}^{2}},
\end{equation}
where $\sigma_{xi}$ and $\sigma_{yi}$ are the $x$ and $y$ standard deviations for the $i$th measurement.  The uncertainties in our data shown in Figures \ref{age_pp} through \ref{mg_fe_pp} are asymmetric.  In computing $\chi^{2}$ we make the following simplification: when the function lies above (below) the data point, we use the upper (lower) uncertainty for $\sigma_{yi}$.  When the function lies to the right (left) of the data point, we use the right (left) uncertainty for $\sigma_{xi}$.  Due to limitation of data points we do not distinguish between High-,
Mid-, and Low-SB when making our fit.  This means that we can only
quantitatively make statements about the differences in the SPPs of our galaxy sample with local group environment
considering $\emph{all}$ fixed points in structural
FP-space.  The linear fit relations we obtain are shown
with a black solid line in each of the panels of Figures \ref{age_pp} through \ref{mg_fe_pp} with the offset fit values and their errors (computed from the variance, formally derived by consideration of propagation of errors, modified by a factor of the reduced-$\chi^{2}$) shown in the bottom right corners.  In each panel we also show a grey dashed line
illustrating the one-to-one relation for comparison.

The results for the ages of BGG, Isolated, and
Satellite galaxies at fixed structure shown in Figure \ref{age_pp} are the
following: BGGs are
older than Isolateds by 0.02 dex (Figure \ref{age_pp}$\mathrm{a}$), Satellites are older
than BGGs by 0.02 dex (Figure \ref{age_pp}$\mathrm{b}$), and Satellites
are older than Isolateds by 0.04 dex (Figure \ref{age_pp}$\mathrm{c}$).
Thus, at fixed structure, Isolated galaxies tend to have the youngest
stellar populations, while BGGs are typically $\sim$5$\%$ older, and
Satellites are the oldest, typically $\sim$10$\%$ older than Isolated galaxies
with the same structure.  These age differences, though small, are
statistically significant.  They are also the largest variations that we
find.

We observe small offsets in [Fe/H] at fixed structure, as shown in
Figure \ref{fe_h_pp}.  BGGs are more Fe-rich than Isolateds by 0.01 dex
(Figure \ref{fe_h_pp}$\mathrm{a}$).  BGGs are also more Fe-rich than
Satellites by 0.02 dex (Figure \ref{fe_h_pp}$\mathrm{b}$).  [Fe/H] is equal between Satellites and Isolateds at fixed
structure (Figure \ref{fe_h_pp}$\mathrm{c}$).  These differences are smaller than
the differences that were seen in age in Figure \ref{age_pp}.

Finally, we find no difference in Mg-enhancement between BGG, Satellite,
and Isolated galaxies at fixed structure (Figure \ref{mg_fe_pp}).

To summarize: the stellar population properties of quiescent early-type
galaxies are well mapped to their structures, with the two most relevant
structural parameters being $\sigma$ and $\Delta I_{e}$; to
zeroth-order, the stellar population properties age, [Fe/H], and [Mg/Fe]
of early-type galaxies are fully captured by their structures, such that
as $\sigma$ of a galaxy increases, age, [Fe/H], and [Mg/Fe] all
increase, while at fixed $\sigma$ as $\Delta I_{e}$ increases, [Fe/H]
increases while age and [Mg/Fe] decrease.

There are slight shifts in the stellar population properties of quiescent early-type galaxies with local group environment on top of the zeroth-order trends with structure.  These are the following: there are modest shifts in age at fixed structure, such that age$_{Sat}$ $>$ age$_{BGG}$ $>$ age$_{Iso}$.  There are $\emph{very}$ slight shifts in Fe-enrichment at fixed structure, such that [Fe/H]$_{BGG}$ $>$ [Fe/H]$_{Sat}$ = [Fe/H]$_{Iso}$.  There are no shifts between the Mg-enhancements of BGG, Isolated, and Satellite galaxies at fixed structure.

\subsection{Star Formation History vs. Velocity Dispersion and Environment}
\begin{figure*}
\begin{center}
\includegraphics[scale=0.65]{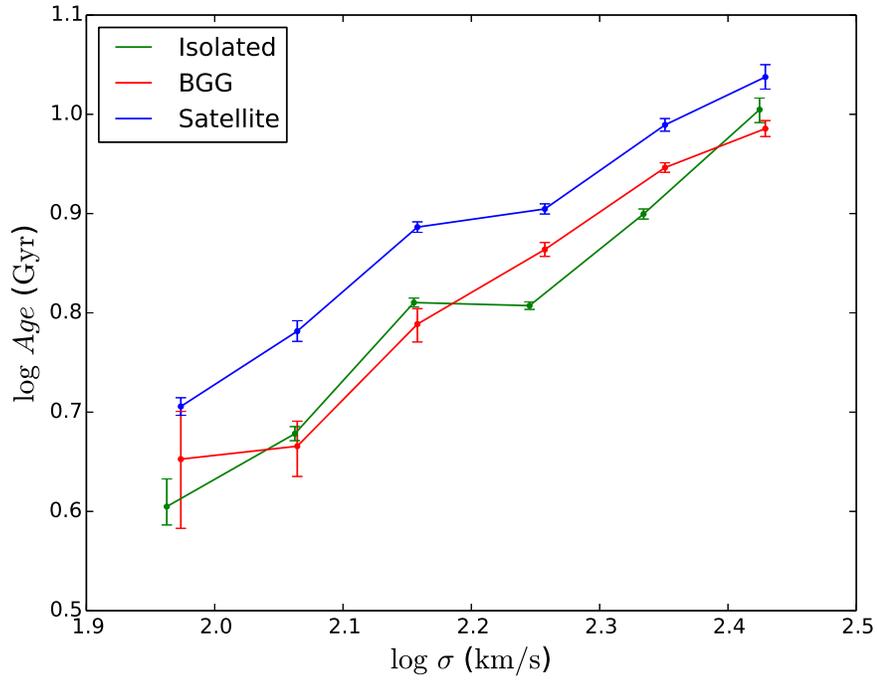}
\caption{The ages of Isolated (\textbf{green}), BGG (\textbf{red}), and Satellite (\textbf{blue}) galaxies in fixed bins of $\sigma$.} \label{sig_age}
\end{center}
\end{figure*}

\begin{figure*}
\begin{center}
\includegraphics[scale=0.65]{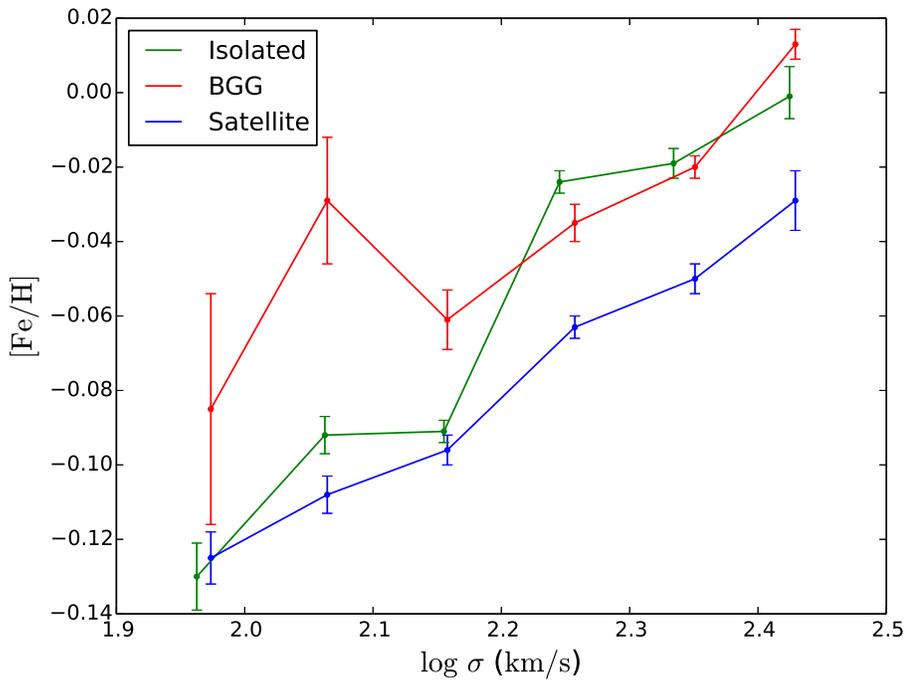}
\caption{The Fe-enrichments of Isolated (\textbf{green}), BGG (\textbf{red}), and Satellite (\textbf{blue}) galaxies in fixed bins of $\sigma$.} \label{sig_fe_h}
\end{center}
\end{figure*}

\begin{figure*}
\begin{center}
\includegraphics[scale=0.6]{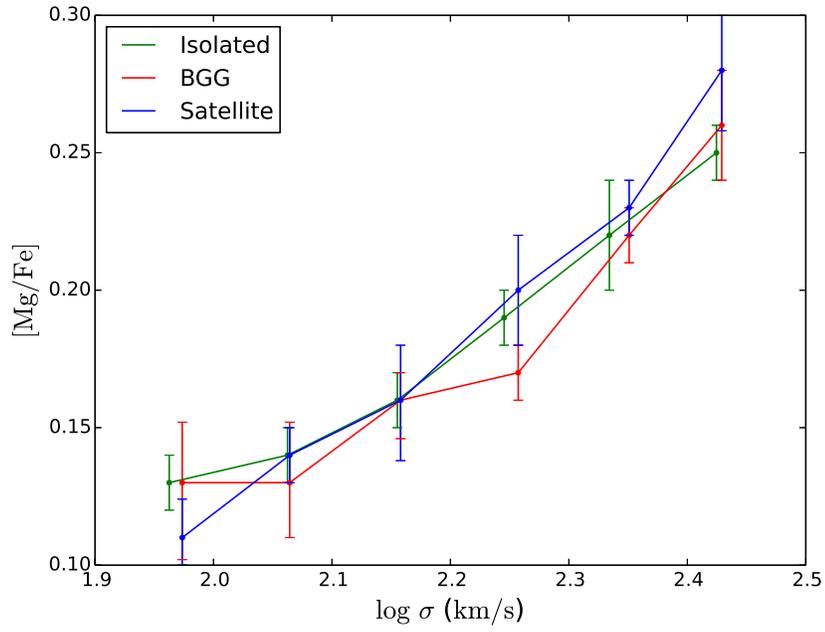}
\caption{The Mg-enhancements of Isolated (\textbf{green}), BGG (\textbf{red}), and Satellite (\textbf{blue}) galaxies in fixed bins of $\sigma$.} \label{sig_mg_fe}
\end{center}
\end{figure*}

Before we discuss the implications of our results at fixed FP-position, we study variation in the SPPs with local group environment at fixed $\sigma$.  This allows us to more clearly place our results at fixed FP-position in the context of previous studies, which have typically studied the SFHs of ETGs at fixed galaxy mass. We measure the ages, [Fe/H], and [Mg/Fe] for the spectra of BGG, Isolated, and Satellite galaxies stacked separately in bins of $\sigma$.  Our results are shown in Figures \ref{sig_age} through \ref{sig_mg_fe}.

At fixed $\sigma$, Satellites are older than BGGs and Isolateds by $\sim$0.1 dex (Figure \ref{sig_age}). BGGs are more Fe-rich than Satellites at fixed $\sigma$ by $\sim$0.04 dex (Figure \ref{sig_fe_h}). The shifts seen in the SPPs between BGG and Satellite galaxies at fixed FP-position are much stronger at fixed $\sigma$. There are no significant shifts in [Mg/Fe] with environment at fixed FP-position or at fixed $\sigma$ (Figure \ref{sig_mg_fe}).

Comparison of the SPPs of BGGs and Satellites with those of Isolateds
in Figures \ref{sig_age} through \ref{sig_mg_fe} is ambiguous because
trends in the SPPs with galaxy structural parameters are not fully
controlled. We can more clearly interpret shifts between the SPPs of
BGGs and Satellites at fixed $\sigma$ because the definitions of these
galaxy subpopulations relative to each other directly indicate how
they will distribute differently across the FP at fixed $\sigma$:
BGGs, by definition, are more massive and brighter -- and so tend to
have higher values of $I_{e}$ -- than Satellites at fixed
$\sigma$. This implies that at fixed $\sigma$, regardless of any true
variation in the SPPs with environment, BGGs tend to be offset to
younger ages and higher [Fe/H] than Satellites. This effect is added
to true variation in the SPPs with environment. Thus, the age and
[Fe/H] shifts between Satellites and BGGs is stronger at fixed
$\sigma$ than it is at fixed FP-position -- in FP-space we control all
these important structural parameters at once and study differences in
SFH with environment that are independent of structure.

\section{Discussion}
The central result of our analysis is that, to zeroth-order, the star
formation histories of quiescent early-type galaxies are observed to be
fully captured by their structures.  The observed strong correlation
between SFH and structure has strong implications for the evolution of
quiescent ETGs after their initial formation and main star-forming
lifetimes. In interpreting our results, we note that our sample
consists of relatively old galaxies that were possibly quenched long ago, which may make environmental effects on
the SFHs difficult to identify, in addition to their intrinsic
subtleness. In considering slight differences in the SFHs with
environment one should also be careful to note that we have not matched the
halo mass distributions between BGGs, Satellites, and Isolateds within
each bin in FP-space. This is most important when
considering BGG and Satellite galaxies, which according to their definitions
cover different ranges of halo masses at fixed galaxy mass.

In the following discussion, we use the term `late-time processes' to describe those evolutionary processes which a galaxy undergoes after its initial formation and main
star-forming lifetime. For quiescent ETGs, late-time processes may
include the quenching of star formation and processes such as dry
merging which occur after quenching.  `Early-time' processes are those
which precede SF quenching.

The observed SFH-structure correlation can result from two possible
scenarios.  Either (1) SFH and structure of a galaxy are both built in
at early times, and are not changed very much by late-time evolutionary
processes, or (2) late-time evolution can significantly change both the
SFHs and structures of quiescent ETGs, but this happens conspiratorially
in a way that preserves the observed SFH-structure correlation.

The possibility that the latter of these scenarios occurs seems unlikely; in particular, because we do not see strong variations in SFH with environment, this scenario would require all late-time environmental processes to follow this same `conspiracy'.  On the other hand, the first scenario is a plausible explanation of our observations.  Our results therefore substantially constrain the effects that late-time evolutionary processes can have on the structures and SFHs of quiescent ETGs.  In the following sections, we discuss the implications of this conclusion for a number of evolutionary processes.

\subsection{Quenching of Infalling Satellites}
\citet{graves10b} suggested that satellite quenching might drive the trends they observed in the SPPs of quiescent ETGs with $\Delta I_{e}$.  Satellite quenching describes processes by which satellites that fall into their host haloes are stripped of cold gas out of which to form stars (e.g. \citealt{gunn72, lea76, gisler76}).  Satellite quenching has been
suggested as an important mechanism causing the observed dependence of
galaxy star formation rate (SFR) and color on small scale environment ($\lesssim$ 1 Mpc), in
which the fraction of red galaxies with low SFR increases with local
density of environment (e.g. Wilman, Zibetti \& Budav\'{a}ri 2010; Tinker, Wetzel \& Conroy 2011). \citet{pasquali10} considered the quenching of star formation of satellite galaxies by strangulation as
they fall into their host haloes to explain the shift they observed of satellites to older ages than centrals at fixed $M_{\star}$ in their SDSS galaxy sample.

If the quenching of
infalling satellites were the dominant mechanism driving the $\Delta
I_{e}$ trends, this
would imply that Low-$\Delta I_{e}$ galaxies tend to be satellite galaxies in
the dark matter haloes of larger host galaxies.  It further implies that satellites offset to
lowest $\Delta I_{e}$ are those which fell into their host haloes at
the earliest times \citep{graves10b}.  The data presented here can rule out the
quenching of infalling satellites as a main mechanism driving the
observed trends with $\Delta I_{e}$.  Considering the bottom panels of
Figure \ref{age}, it is clear that the Low-$\Delta I_{e}$ galaxies include BGG, Isolated, and Satellite galaxies in similar proportion to what
is observed for High- and Mid-$\Delta I_{e}$ galaxies.  Thus, the Low-$\Delta
I_{e}$ population is not especially dominated by
Satellites.

Although satellite quenching is not the main mechanism
driving the observed trends in the SPPs with
$\Delta I_{e}$, our results are still consistent with the presence of
some quenching of infalling satellites.  In particular, our observed
slight shifts in the ages of Satellites above BGGs and Isolateds at
fixed structure might be due in part to the quenching of these
Satellites as they fall into their host haloes.

If the
quenching of infalling Satellites is indeed the mechanism causing the observed age offset,
then we might expect Satellites to also be
offset to lower [Fe/H] and higher [Mg/Fe]
than Isolated galaxies, consistent with a mechanism that truncates the
star formation of Satellite galaxies more efficiently than that of
Isolated galaxies.  We do not, however, observe either of these shifts.
In fact, we observe that [Fe/H]$_{Sat}$ = [Fe/H]$_{Iso}$, and
that [Mg/Fe]$_{Sat}$ = [Mg/Fe]$_{Iso}$.

This indicates that if
satellite quenching is the mechanism causing the shift of Satellites to
older ages than Isolateds and BGGs, this must only be a weak and late
effect that takes place after a Satellite galaxy has been forming
Fe-peak elements for a significant period of time.  This is also
consistent with satellite quenching which occurs over a long time
period, rather than abruptly.  This is indeed suggested by the results
of \citet{wetzel13}, who, using galaxy group/cluster catalogs from
the SDSS DR7 together with a cosmological N-body simulation, find that
there is a long time delay (2--4 Gyr) in which, after a satellite galaxy
falls into its host dark matter halo, satellite SFRs evolve unaffected
before an ensuing period of rapid quenching with a characteristic
time-scale $<$0.8 Gyr.

The presence of slow, weak satellite quenching in our sample is thus consistent with the observed trends in age, [Fe/H], and [Mg/Fe] with local group environment.  That satellite quenching can have only a subtle effect on galaxies in our sample, in a way consistent with the zeroth-order preservation of their SFHs/structures, implies that the mechanisms responsible for quenching Satellites after they fall into their host dark matter haloes cannot significantly alter the structures of Satellite galaxies.  This is supported by the results of \citet{vandenbosch08}, who, using a large galaxy group catalog constructed from SDSS, suggest that `strangulation' (Larson, Tinsley \& Caldwell 1980; Balogh, Navarro \& Morris 2000) is the main mechanism turning satellite galaxies red, dominating over mechanisms which operate only in very massive haloes, such as `ram-pressure stripping' (e.g. \citealt{gunn72}; Quilis, Moore \& Bower 2000; \citealt{hester06}) and `harrasment' (\citealt{farouki81, moore96}), the latter of which can have a significant impact on galaxy morphology.

\subsection{Cooling Flows}
The efficiency of cooling flows in our sample might be
expected to differ with local group environment, and could impact the observed SFHs.  Cooling flows significantly affect only the galaxy at
the bottom of the gravitational potential well of its halo, and so should only be relevant for BGG and Isolated
galaxies.

Late-time star formation has commonly been
observed in `cooling core' BCGs with low central gas entropies,
identified by their X-ray luminosities and excess IR and UV emission
(e.g. \citealt{egami06, donahue07, quillen08, odea08, wang10, hoffer12}).  In analogy to the BCGs in clusters, we might
expect to see subtle shifts in BGGs to younger ages than Isolateds, as a
result of more efficient gas cooling flows powering residual late-time
star formation for central BGG galaxies.  This is consistent with
observations which find galaxy clusters with the most massive cooling
flows to have significantly closer nearest neighbors than the typical
cluster (Loken, Melott \& Miller 1999).  This could be driven by a positive
correlation between cluster nearest-neighbor overdensity and central gas
density \citep{miller99}.  By definition, BGGs should reside in regions
of higher nearest-neighbor overdensity than Isolateds and have higher
central gas density than Isolateds, and could therefore have more
efficient cooling flows.  We note that we have eliminated ongoing star
formation from our sample, but this only eliminates continuous residual
late-time star formation, and any residual star-formation that shut down
recently will still affect our results.

We do not, however, observe BGGs to be offset to younger ages at fixed structure than Isolateds and, in fact, the slight shift we do see is an offset of BGGs to slightly $\emph{older}$ ages than Isolateds at fixed structure.  Not only do we not see evidence for significant late-time residual star formation from more efficient cooling flows for BGGs in our sample, but the slight shift is in the opposite direction from what one might expect if this were the case.  We conclude that cooling flows do not have a major impact on the integrated SFHs of BGGs in our sample.

\subsection{Quenching in Massive Haloes}
The observed small offset of BGGs to slightly
older ages than Isolateds could be due to more efficient late-time
residual star formation for Isolateds.  In this case, a possible mechanism causing this
could be quenching by massive haloes.  This occurs when the dark matter
halo of a galaxy passes a critical mass threshold and accreting gas is
shock heated upon infall, no longer cooling efficiently (e.g. \citealt{birnboim03, keres05, cattaneo08}).
The interaction of the accreting gas with the rapidly expanding shock
has been shown to be enough for long-term quenching in haloes of
10$^{12}$ to 10$^{13}M_{\odot}$ (Birnboim, Dekel \& Neistein 2007).  It has also been shown that long-term quenching in massive ($\gtrsim$
7$\times$10$^{12}$M$_{\odot}$) haloes can result if shock heating is
supplemented by continued cosmological gas accretion which delivers
gravitational energy to the inner halo hot gas \citep{dekel08}.

At fixed structure of the central galaxy, we expect BGGs to reside in
more massive dark matter haloes than Isolateds because they are defined
as having a larger total group mass.  It is possible that gas accretion
onto Isolateds is more efficient, while shock heating slows or quenches
star formation for BGGs in more massive haloes.  This could cause our
observed slight shift of BGGs to older ages than Isolateds at fixed
structure.  We note that this mechanism is thought only to be effective
when the dark matter halo of a galaxy passes beyond a critical mass
threshold $M_{crit}$.  In order to test this further, it would be
interesting in later similar analyses to systematically determine how
the SPPs of quiescent ETGs scale with local group environment at fixed
structure, distinguishing by galaxy halo mass.

Besides a slight age shift, we also see a very slight shift of BGGs to
higher [Fe/H] than Isolateds and Satellites.  In a way that is
consistent with the critical halo mass quenching arguments above, the
observed shift in [Fe/H] of BGGs can also be explained by their larger
mass haloes at fixed structure.  If, at fixed structure, BGGs have
more massive haloes than Isolateds, the [Fe/H] offset of BGGs could
result from BGGs being slightly less susceptible to SN feedback,
retaining a larger fraction of their metals in the deeper gravitational potential
wells in which they reside.

The effects of differences in the halo masses of BGGs and Isolateds on their SFHs and structures cannot be large, as implied by the only slight shifts seen in the SFHs of galaxies in our sample with environment, in a way that is consistent with the zeroth-order preservation of structure and SFH.

\subsection{Dry Merging}
We would expect the rate of dissipationless (`dry')
merging in our galaxy sample to depend on environment.  In particular,
we might expect the rate of dry minor merging to be greater for BGGs.
This is consistent with the results of \citet{bernardi09} who suggests
that more dry minor merging in the formation histories of BCGs
(analogous to our BGGs) causes their steeper $R_{e}$-$L$ relation over
other ETGs in the SDSS, MaxBCG \citep{koester07}, and C4
\citep{miller05} catalogues.  \citet{ruszkowski09}, performing
high-resolution cosmological simulations, also find BCGs to evolve away
from the Kormendy relation for smaller mass galaxies and to show
increased dark matter-to-stellar mass ratios, due to a larger number of
dry mergers in their formation histories.  Because dry merger rates in
our sample are expected to vary with environment, our observation of the
zeroth-order preservation of SFH-structure constrains the degree to
which dry merging can alter galaxy structure without `equally' altering
the SPPs.

Naab, Johansson \& Ostriker (2009) and \citet{bezanson09} present simple scaling arguments
to discuss the effects of dry mergers on ETG structure.  These scaling
arguments have agreed reasonably well with the results of simulations
(e.g. Oser et al. 2012; Boylan-Kolchin, Ma \& Quataert 2006). Using
the virial theorem, and assuming energy conservation between the
compact initial stellar system and the final system after accretion of
additional stellar systems, \citet{naab09} derives the following
relation between the final and initial mean square speeds of stars:
\begin{equation}
\frac{\langle v_{f}^{2} \rangle}{\langle v_{i}^{2} \rangle} =
\frac{1 + \eta \epsilon}{1 + \eta},
\label{eq_rat_v}
\end{equation}
and a similar relation between the final and initial gravitational
radii:
\begin{equation}
\frac{r_{g,f}}{r_{g,i}} = \frac{\left( 1 + \eta \right)^{2}}{1 + \eta
  \epsilon},
\label{eq_ra_t}
\end{equation}
where $\eta$ is the ratio of the total mass of accreted systems to the
initial mass ($\eta$ = $M_{a}/M_{i}$), and $\epsilon$ is the ratio of the average of the mean
square speeds of accreted systems and the initial mean square speed
($\epsilon$ = $\langle v_{a}^{2} \rangle / \langle v_{i}^{2}
\rangle$).

These scaling arguments predict that the remnant $R$ in a dry major merger doubles for a factor-of-2 increase in mass while its $\sigma$ is left unchanged.  Referring to
Equation \ref{FP_eq}, if dry major mergers increase $R$ linearly with
mass, then they would leave the structural parameters relevant to the
SFHs of galaxies in our sample relatively unchanged.  Because they also
occur between galaxies of similar structure (and SFH), dry major mergers
are consistent with our observation of the zeroth-order preservation of
SFH-structure.

The \citet{naab09} and \citet{bezanson09} scaling arguments predict that
for a factor-of-2 increase in mass due to dry minor merging, $R$
increases by a factor of 4 and $\sigma$ decreases by a factor of
$\sqrt{2}$.  Again referring to Equation \ref{FP_eq}, we find that,
according to these scaling arguments, enough dry minor merging can
significantly alter the relevant structural parameters of galaxies in
our sample (decreasing their $\sigma$ and shifting them to lower
$I_{e}$).  Because we expect dry minor mergers to leave the SPPs of
the most massive progenitor relatively unchanged, a large amount of dry minor
merging would `wash out' the observed SFH-structure correlations, in disagreement with the data.  This
constrains the rates of dry minor mergers in our sample and the degree
to which they can alter galaxy structure.

Limitations on dry minor merger rates and structural effects provided by
our observations could possibly conflict with studies which favor dry
minor merging as a mechanism to explain the observed size growth (up to
a factor of $\sim$5 since a redshift of $\sim$2) of elliptical galaxies
(e.g. \citealt{bezanson09, hopkins10, oser10, oser12, vandokkum10}; Trujillo, Ferreras \& de la Rosa 2011; \citealt{hilz12}; Hilz, Naab \& Ostriker 2013; \citealt{bernardi09, vanderwel09, naab09, oogi13,
  laporte13}).  Indeed, as our observations might suggest, recent
studies have found that the observed size growth of ETGs cannot be fully
attributed to dry minor mergers (Nipoti, Treu \& Bolton 2009a; \citealt{nipoti09b, nipoti12}; Cimatti, Nipoti \& Cassata 2012).  \citet{nipoti09b}, using N-body simulations,
found dry major and minor mergers together to increase the half-light
radii and velocity dispersions of ETGs with $M_{\star}$ as $R_{e}$
$\propto$ $M_{\star}^{1.09 \pm 0.29}$ and $\sigma$ $\propto$
$M_{\star}^{0.07 \pm 0.11}$.  Based on their simulations, the authors
determined that if high-redshift ETGs are indeed as dense as estimated,
they cannot have evolved into present-day ETGs via dry merging alone.
In particular, the authors found that present-day ETGs cannot have
assembled more than $\sim$45$\%$ of their stellar mass via dry mergers,
and even this upper limit requires extreme fine tuning (see also
\citealt{nipoti09a, nipoti12}).

We note that on average the SDSS fiber only samples the inner
0.78$R_{e}$ of galaxies in our sample, and only samples $\leq$0.5$R_{e}$
for $\sim$20$\%$ of our sample.  Dry minor mergers are expected to
accrete material at large radii in a galaxy, forming extended stellar
envelopes of low-density material (e.g. \citealt{naab07, naab09, oser10, oser12}; Farouki, Shapiro \& Duncan 1983; \citealt{villumsen83}), but may have a minimal effect on the
central regions of a galaxy (e.g. \citealt{naab07, bezanson09, hopkins09}).  In particular, \citet{hopkins09} found that although the
effective stellar mass surface densities within $R_{e}$ ($\Sigma_{e}
\equiv 1/2M_{\star}/ \left( \pi R_{e}^{2} \right)$) of local massive
ellipticals ($>$10$^{11} M_{\odot}$) is much smaller than those of their
high-redshift counterparts, the physical stellar surface densities at
the observed radii $\sim$1--5 kpc are comparable between low- and
high-redshift massive ellipticals.  Evidence was found for a picture in
which the entire population of high-redshift red galaxies are the
progenitors of the high-density cores of present-day massive
ellipticals, whose growth is primarily through dissipationless minor
mergers.  \citet{bezanson09} found similar results.  Thus, although it
does not entirely lift any constraint our observations have on the rates and
effects of dry minor merging allowed for in our sample (in particular,
the full $R_{e}$ is sampled by the SDSS aperture for $\sim$50$\%$ of
galaxies in our sample), any attempt to place a limit on the effects of
dry minor merging in our sample must consider the effect of finite SDSS
aperture size.

It would be interesting in future studies to perform numerical simulations to quantify the amount of dry merging allowed in our galaxy sample before the SFH-structure correlation we observe is obscured, taking into account the effect of finite SDSS aperture size.  Such a study could help place quantitative constraints on structural evolution of ETGs due to dry merging and the contribution of dry merging to the observed size growth of ETGs.

\section{Conclusion}
In this analysis we used very high S/N, stacked spectra
of $\sim$29,000 SDSS quiescent early-type galaxies to study variations
in the stellar population properties age, [Fe/H], and [Mg/Fe] with local
group environment (BGG, Isolated, and Satellite) at fixed position along
and through the Fundamental Plane.  By fixing galaxies along the
Fundamental Plane parameters $\sigma$ and $\Delta I_{e}$ which were
previously shown to be well-correlated to the star formation histories
of early-type galaxies (\citealt{graves10b, springob12}), we were able
to study variations in the stellar population properties of early-type
galaxies due solely to environment.

We find the following results for the stellar populations of quiescent early-type galaxies:
\begin{enumerate}
\item We confirm the trends in the stellar population properties with galaxy structure seen by \citet{graves10b} and \citet{springob12}: the ages, [Fe/H], and [Mg/Fe] of our galaxy sample all increase with $\sigma$.  Along decreasing $\Delta I_{e}$, galaxy age and [Mg/Fe] increase while [Fe/H] decreases.

\item Our central result is that, to zeroth-order, the star formation
  histories of our early-type galaxy sample are fully captured by the
  structural parameters $\sigma$ and $\Delta I_{e}$, and any
  differences in the star formation histories with environment at fixed
  structure are only slight.  The SFH-structure correlation we observe constrains the degree to which late-time evolutionary processes can alter the SFHs/structures of early-type galaxies in our sample.

\item On top of the zeroth-order SFH-structure correlation, there are slight variations in the SFHs of early-type galaxies in our sample with environment: Isolated galaxies have the youngest ages, while BGGs are 0.02 dex older, and Satellites have the oldest stellar populations, 0.04 dex older than Isolateds.  BGGs are found to have the highest Fe-enrichments, 0.01 dex higher than Isolateds and 0.02 dex higher than Satellites.  Satellites and Isolateds have equal Fe-enrichments.  There are no differences in Mg-enhancement between BGG, Isolated, and Satellite galaxies.
\end{enumerate}

Quiescent early-type galaxies in our sample obey a SFH-structure correlation that is determined early-on and preserved throughout late-time evolution.  On top of this correlation there are only slight trends in SFH with environment.  Although satellite
quenching is found not to be the main mechanism causing the truncation sequence
observed along the $\Delta I_{e}$ dimension of FP-space, as proposed by \citet{graves10b}, our observation that Satellites are
slightly offset to older ages than BGGs and Isolateds is consistent
with a weak, slow satellite quenching process.  Although we do not see
any slight offset of BGGs to younger ages than Isolateds, as one might
expect from cooling flows, this could be due to quenching in the
more massive haloes in which BGGs reside.  The strong SFH-structure correlation we observe may well be inconsistent with a large amount of dry minor merging in our sample.  Future numerical simulations are needed to place quantitative constraints on the degree to which dry minor merging can affect our galaxy sample.

\section*{Acknowledgements}

The authors would like to thank Anthony Paredes, Garrett Keating, Mariska Kriek, and Dan Kasen for valuable
discussions. The authors would also like to thank Charlie Conroy and an anonymous referee for helpful suggestions.  G. G. acknowledges support from the Miller
Institute for Basic Research in Science during the duration of this
project. 

Funding for the creation and distribution of the SDSS Archive has been
provided by the Alfred P. Sloan Foundation, the Participating
Institutions, the National Aeronautics and Space Administration, the
National Science Foundation, the US Department of Energy, the Japanese
Monbukagakusho, and the Max-Planck Society. The SDSS Web site is
http://www.sdss.org/.

The SDSS is managed by the Astrophysical Research Consortium (ARC) for
the Participating Institutions. The Participating Institutions are the
University of Chicago, Fermilab, the Institute for Advanced Study, the
Japan Participation Group, the Johns Hopkins University, the Korean
Scientist Group, Los Alamos National Laboratory, the
Max-Planck-Institute for Astronomy (MPIA), the Max-Planck-Institute
for Astrophysics (MPA), New Mexico State University, University of
Pittsburgh, University of Portsmouth, Princeton University, the United
States Naval Observatory, and the University of Washington.

\clearpage

\end{document}